\documentclass[a4paper,12pt]{article}
\pdfoutput=1
\usepackage{epsfig}
\usepackage[titletoc,title]{appendix}
\usepackage{epstopdf}\usepackage[sumlimits,intlimits,namelimits]{amsmath} 
\usepackage{dsfont}
\usepackage{slashed}

\usepackage[english]{babel}

\usepackage{graphicx}

\usepackage{geometry}
\makeatletter
\def\fmslash{\@ifnextchar[{\fmsl@sh}{\fmsl@sh[0mu]}}
\def\fmsl@sh[#1]#2{%
  \mathchoice
    {\@fmsl@sh\displaystyle{#1}{#2}}%
    {\@fmsl@sh\textstyle{#1}{#2}}%
    {\@fmsl@sh\scriptstyle{#1}{#2}}%
    {\@fmsl@sh\scriptscriptstyle{#1}{#2}}}
\def\@fmsl@sh#1#2#3{\m@th\ooalign{$\hfil#1\mkern#2/\hfil$\crcr$#1#3$}}
\makeatother

\newcommand{\ep}{\epsilon}

\newcommand{\ice}[1]{\relax}

\usepackage{a4wide,amsmath,graphicx,calc}
\newlength{\figh}
\numberwithin{equation}{section}
\begin{document}
\begin{titlepage}
\begin{flushright}
SI-HEP-2018-13 \\
QFET-2018-07\\
MITP/18-035
\end{flushright}

\begin{center}
{\Large\bf 
  {\boldmath $B^0$-$\bar{B}^0$ \unboldmath} mixing:\\
  matching to HQET at NNLO}
\end{center}

\begin{center}
\textsc{Andrey G.~Grozin}$^{a,b,c,d}$,\\
\textsc{Thomas Mannel}$^b$ and \textsc{Alexei A.~Pivovarov}$^b$\\[0.1cm]
$^a$ \textsf{PRISMA Cluster of Excellence, Johannes Gutenberg Universit\"at,}\\
\textsf{Staudingerweg 9, 55128 Mainz, Germany}\\
$^b$ \textsf{Theoretische Elementarteilchenphysik, Naturwissenschaftlich-Technische Fakult\"at,}\\
\textsf{Universit\"at Siegen, Walter-Flex-Str.~3, 57068 Siegen, Germany}\\
$^c$ \textsf{Budker Institute of Nuclear Physics SB RAS, Lavrentyev st.~11, Novosibirsk 630090, Russia}\\
$^d$ \textsf{Novosibirsk State University, Pirogova st.~2, Novosibirsk 630090, Russia}
\end{center}

\begin{abstract}
\noindent
We compute perturbative matching coefficients
to the Heavy Quark Effective Theory representation
for the QCD effective local 
$\Delta B=2$ Hamiltonian
that determines the mass difference in $B^0-\bar{B}^0$
system of states.
We report on the results at NNLO in the strong coupling constant
for matching coefficients of two physical operators in HQET.
Our results provide firm confirmation that the recent NLO 
sum
rules analysis of the bag parameter $B_q$ 
is stable with regard of inclusion of higher order
radiative corrections. 
As a by-product of our calculation we give a fully analytical solution
for the one loop QCD-to-HQET
matching problem: we present the explicit formulas for the
renormalization of four quark
operators of the full bases in both QCD and HQET and the expressions
for matching coefficients in a closed form. 
\end{abstract}


\end{titlepage}

\pagenumbering{arabic}

\section{Introduction}
\label{Sect:Intro}
The accuracy of theoretical predictions for the neutral meson mixing 
within the Standard Model (SM) has been steadily improving during last
years~\cite{Lenz:2006hd,Lenz,Nierste}. The main reason for this progress
is due to better numerical precision achieved for the numerical values
for hadronic matrix elements in the
lattice simulations~(e.g.~\cite{Aoki:2016frl,Aoki:2014nga}).
Nevertheless, recently some new results
with a competitive level of accuracy
appeared in the domain of analytical computation based on sum rules
approach~\cite{Grozin:2016uqy,Kirk:2017juj}.
The latter calculation has become possible due to the
technical advance of integral computation
at three loops~\cite{GL:09}.
At present, the high precision of experimental material 
provides for good opportunities 
for searches of physics beyond SM~\cite{Dowdall:2014qka},
and
these new theoretical predictions for the mixing within SM are
important since the search for new physics
depends heavily on the accurate knowledge of numerical values of low
energy parameters~\cite{Carrasco:2013zta,Bazavov:2016nty}. 
The mixing of the neutral 
$B$ mesons is dominated by the top-quark contribution and hence 
looks directly to the ultraviolet (UV) new physics having long 
distance effects under control.
The concept of effective theories allows for getting better precision
for the theoretical predictions due to
separation of scales~\cite{Buras:1990fn,BBL:96,Beneke:1998sy}
that essentially
improves on old electroweak
results~\cite{Inami:1980fz}. 
Clearly, the most crucial point of such an improvement 
is perturbation theory (PT) 
corrections at the lowest scales involved in the analysis.
This motivates our research -- we compute next-to-next-to leading
order (NNLO) corrections
to the matching of QCD to heavy quark effective theory (HQET) 
that is applicable at the scales of order few $\Lambda_{\text{QCD}}$.
Some partial results have been already published
earlier~\cite{Grozin:2017uto}.

The paper is organized as follows.
In the next section
we briefly introduce notation just for the paper to be self-contained
(the details can be found in~\cite{Grozin:2017uto}) 
and give our main results. In Sect.~\ref{Sect:comp} we describe the
technique of the computation.  In Sect.~\ref{Sect:pheno}
we briefly discuss implications of our calculation for phenomenology.
In summary section we give our conclusions.

\section{Operators and Wilson coefficients}
\label{Sect:basis}
In the SM the $B^0$-$\bar{B}^0$ transition is described 
by a nonlocal expression of
quark scattering at loop level. 
The most important corrections to the leading  electroweak term  
are the contributions of
strong interactions. They can be computed within QCD perturbatively
as the relevant scale is of order the meson mass $m_B$ and is much larger
than QCD infrared scale $\Lambda_{\text{QCD}}$.

In this section
we repeat some formulas from~\cite{Grozin:2017uto}
to introduce notation. More details are given
in~\cite{Grozin:2017uto}.

\subsection{QCD: below $m_W$}
A relevant scale for the description of $B^0$-$\bar{B}^0$ 
physics is around the
$B$-meson mass $m_B$ that is kinematically
saturated by the $b$-quark mass 
$m_b$.
After integrating out particles of the SM that are heavier
than $b$-quark the effective Hamiltonian
for the process of $B^0$-$\bar{B}^0$ mixing reads
\begin{equation}
{\cal H}=C(m_W, m_t,\mu, \alpha_s(\mu))Q(\mu)
\label{eq:effH}
\end{equation}
with 
\begin{eqnarray}
Q(\mu)=\bar q_L \gamma_\alpha b\bar q_L \gamma_\alpha b(\mu)\, .
\label{eq:Q}
\end{eqnarray}
The quantity $Q(\mu)$ is a local renormalized operator with
$\Delta
B=2$.
One can choose a
low normalization point $\mu\sim m_b$ for the operator $Q$
while large
logarithms of the scales ratio, $\ln(m_W/m_b)$,
in the Wilson coefficient
$C(m_W, m_t,\mu,\alpha_s(\mu))$ can be regularly resummed with
renormalization group (RG) 
techniques. 
At present the coefficient $C(m_W, m_t,\mu,\alpha_s(\mu))$
is known at the next-to-leading order (NLO) of the expansion in the strong
coupling constant that gives the accuracy of a few 
percent~\cite{Buras:1990fn,BBL:96,Beneke:1998sy}.
The renormalization properties of the operator $Q$
are a bit peculiar but well understood
and intensively discussed in the literature.
The point is that the calculation are usually and dominantly
performed in dimensional regularization and
one has to close up the algebra of Dirac gamma matrices which is
infinitely dimensional in general $d$-dimensional space. This is an
outstanding problem and has been much
discussed the literature~\cite{tHooft:1972tcz}.
Dimensional reduction has been used for the 
computation of corrections to $\Delta F=1$
Hamiltonian~\cite{Altarelli:1980fi}.
A naive dimensional regularization
prescription (NDR) is introduced in~\cite{Chanowitz:1979zu}.
As a result of using dimensional regularization for the integral
evaluation one 
requires a special treatment of Dirac structure and, eventually, the
extension of the operator basis.
For the four quark operator in question, i.e. $Q(\mu)$ in eq.~(\ref{eq:Q}),
the procedure was discussed by  
Buras {\it et al.}~\cite{Buras:1989xd}.
The method has been further analyzed
in~\cite{Dugan:1990df,Herrlich:1994kh}.
Note that the anomalous dimensions of baryon
or three quark operators have been computed earlier within  
a similar approach~\cite{Pivovarov:1988gt,Pivovarov:1991nk}. A clear
presentation of the techniques is given in~\cite{Chetyrkin:1997gb}.

Thus, the effective Hamiltonian in eq.~(\ref{eq:effH})
should,in fact, contain additional evanescent
operators and should explicitly read as
\begin{equation}
{\cal H}=C(m_W, m_t,\mu, \alpha_s(\mu))Q(\mu)+C_E E
\label{eq:EffHevan}
\end{equation}
where $E$ is a general notation for a string of evanescent
operators in QCD. Their matrix elements vanish but their presence
in the basis 
influences the renormalization pattern and, therefore, evolution of
physical operators $Q(\mu)$.   
The renormalized operator $Q(\mu)$
depends on the choice of the evanescent ones.
By choosing $\mathrm{E}' = \mathrm{E} + a \epsilon \mathrm{Q}$
one obtains a different renormalized physical
operator $Q'(\mu)$.
At the one loop level one gets the relation between the two
\begin{equation}
Q'(\mu) = \left[1 - a z_{QE} \frac{\alpha_s(\mu)}{4\pi}\right]
Q(\mu)\, .
\label{PhysEva}
\end{equation}
Physical predictions stay independent of the choice of evanescent
operators. We discuss how it is achieved later in the text.

\subsection{HQET: below $m_b$}
Since
the scale $m_b$
is still QCD perturbative, $m_b\gg \Lambda_{\text{QCD}}$,
one can go further low in scales and remove the explicit dependence on $m_b$
from the matrix element or the mixing amplitude at low energy.
This is achieved by using HQET~\cite{N:94,MW:00,G:04}.

The low scale for the operators involved in the mixing
used to be necessary for the lattice computation,
however
at present there is sufficient power for lattice simulations directly
at the scale $m_b$.
In a computational framework
within analytical methods one matches the theory of QCD on to
HQET where considerable technical simplifications occur for subsequent
computation of sum rules. Thus the matching is an unavoidable part
of the whole computation.
Though there is an approach based on calculation at
$m_b$~\cite{Korner:2003zk}. The results need update as the definition
of the operator has been different.

The heavy quark expansion (HQE)
of the operator $Q(\mu)$ goes
\begin{equation}
Q(\mu) = 2 \sum_{i=1}^2 C_i(\mu) {\tilde O}_i(\mu) +
\mathcal{O}\left(\frac{\Lambda_{\text{QCD}}}{m_b}\right)\, .
\label{eq:match}
\end{equation}
with ${\tilde O}_1(\mu)={O}_l(\mu),~{\tilde O}_2(\mu)={O}_s(\mu)$.
The HQET operators ${O}_{l,s}(\mu)$
are defined as 
\begin{eqnarray}\label{eq:original_matching}
O_l=(\bar q_L \gamma_\mu h_+)( \bar q_L \gamma_\mu h_- ),\quad
O_s=(\bar q_L h_+)( \bar q_L h_-)\, .
\end{eqnarray}
The field $h_+$ annihilates the heavy quark in HQET
(moving
with the four velocity $v$),
and $h_-$ creates the heavy antiquark (again moving with 
the four velocity $v$), 
which is a completely separate particle in HQET framework.

In HQET one can define 
its own set of physical and evanescent operators
(see,~\cite{Grozin:2017uto}).
The general basis is 
\begin{eqnarray}
O_n=(\bar q \gamma_\perp^n h_+)( \bar q \gamma_\perp^n h_- ),\quad
O_n^\prime=(\bar q_i \gamma_\perp^n h_+^j)
( \bar q_j \gamma_\perp^n h_-^i ),\quad
\end{eqnarray}
and $q$ is a light fermion which is usually a chiral one, $q\equiv
q_L$.
A choice for a basis in HQET 
is an antisymmetrized product of transverse gamma's,
\begin{eqnarray}
\gamma_\perp^\mu=\gamma^\mu-v^\mu\slashed{v}\, .
\end{eqnarray}
Then $\gamma_\perp^n$ is a notation for
an antisymmetrized product of $n$ transverse 
gamma matrices. We  sometimes call the number $n$ a rank of the
product and, therefore, of the corresponding operator.

For our two loop calculation it is more convenient to change the
operator basis in the physical sector 
from the standard $\{O_l,O_s\}$~\cite{Grozin:2017uto}
to  $\{O_l,O_p\}$ with 
\begin{eqnarray}
O_p=O_s + \frac{1}{4}O_l
\end{eqnarray}
The operators $\{O_p,O_l\}$ do not mix under renormalization
at the one-loop level~\cite{CFG:96}.
The expression of operators  $\{O_l,O_p\}$
through the basis operators $\{O_n,O_n'\}$
is
\begin{eqnarray}
O_l=O_1-O_0,\quad\quad O_p= \frac{3}{4}O_0 + \frac{1}{4}O_1
\end{eqnarray}

The matching pattern of the QCD operator $Q(\mu)$ at $\mu=m_b$ then
reads
\begin{equation}
Q(m_b) = 2\{  C_l(m_b) O_l(m_b)+  C_p(m_b) O_p(m_b)  + \ldots \}\, 
\label{eq:match1}
\end{equation}
where dots denote the contribution of evanescent operators.

We define a PT expansion of the matching
coefficients $C_{l,p}(m_b)$
as
\begin{eqnarray}  
  C_l(m_b) &=& 1 + \frac{\alpha_s(m_b)}{4\pi} C_l^{(1)}
          + \left(\frac{\alpha_s(m_b)}{4\pi}\right)^2 C_l^{(2)}\,,\nonumber\\
  C_p(m_b) &=& \frac{\alpha_s(m_b)}{4\pi} C_p^{(1)}
          + \left(\frac{\alpha_s(m_b)}{4\pi}\right)^2 C_p^{(2)}\, .
          \label{eq:coefexp}
\end{eqnarray}
Both $\{O_l,O_s\}$ and  $\{O_l,O_p\}$ bases are convenient
since at the leading order there is a
single operator $O_l$ in the matching relation and the other operator
($O_s$ or $O_p$) first appears at NLO.

In HQET limit the theory has $n_l$ massless
flavors and the coupling constant in
eq.~(\ref{eq:coefexp})
is
defined
accordingly as $\alpha_s^{(n_l)}$.

At NLO the values of the matching coefficients in $\{O_l,O_p\}$ basis are~\cite{FHH:91,CFG:96,B:96}
\begin{equation}
C_l^{(1)}(m_b) = - \frac{(N-1) (7N+15)}{2 N},\quad
C_p^{(1)}(m_b) = -2(N+1)\,.
\end{equation}
where $N$ is a number of colors for the $SU(N)$ gauge group.
Note the difference with the  $\{O_l,O_s\}$ basis that is traditionally
used at NLO in the literature
\begin{equation}
C_l^{(1)}(\text{ls-basis}) = \frac{-8N^2-9N+15}{2N}\, .
\end{equation}

In the present paper we have computed
the NNLO contributions to the coefficients $\{C_l,C_p\}$
which is the main result of the paper.

The coefficient  $C_p^{(2)}(m_b)$ reads
\begin{align}
C_p^{(2)} ={}& (N+1) \biggl[
\frac{38}{9} n_l
- \frac{8}{3} I_0
- \frac{2}{9} \frac{4 N^2 + 9 N - 29}{N} \pi^2
\nonumber\\
&{} - \frac{686 N^3 - 563 N^2 + 1599 N + 18}{36 N^2}
\biggr]\,.
\label{eq:cp}
\end{align}
Here $I_0$ is one of the master integrals of the computation that reads
\[
I_0  = \pi^2\log(2) - \frac{3}{2}\zeta(3)= 5.038\ldots
\]
Note that the PT expansion in HQET goes over the $n_l$-flavored
coupling constant since there are just  $n_l$ flavors in the
low energy theory.
In QCD we have in addition
an active heavy quark $b$ and, therefore, $n_l+1$ flavors.

For the number of colors $N=3$ and the number of massless flavors
$n_l= 4$, the numerical values of the expansion coefficients are
 \[
C_p\sim   \{0,-8,-311.166\}\, .
 \]
One sees that the convergence of PT series for quantity $C_p(m_b)$
in the renormalized
coupling constant $\alpha_s(m_b)$ with $n_l=4$ is marginal.
 
The coefficient  $C_l^{(2)}$ is
\begin{align}
C_l^{(2)} ={}& (N-1) \biggl[
\left( \frac{N+3}{3 N} \pi^2 + \frac{123 N + 211}{24 N} \right) n_l
- 2 \frac{N^2 + N + 1}{N^2} I_0
\nonumber\\
&{} - 2 (N-1) \frac{N^2 + 2 N + 2}{N^2} \zeta(3)
- \frac{43 N^3 + 111 N^2 - 111 N - 275}{48 N^2} \pi^2
\nonumber\\
&{} - \frac{13518 N^3 + 8456 N^2 - 7981 N + 35037}{576 N^2}
\biggr]\,.
\label{eq:cl}
\end{align}
For $N=3$ and $n_l= 4$ the numerical values of the coefficients of the
consecutive powers of the coupling constant are
\[
C_l\sim  \{1, -12, -175.559\ldots\}\,  .
\]
The coefficient of the $n_l$ structure
is different from the one given in
our early paper~\cite{Grozin:2017uto} for two reasons. First, the mixing
of operators $\{O_l,O_s\}$ has not been accounted for,
and second, the expansion of
$\Gamma(\epsilon)$ at NLO has not been done to a necessary order in
$\varepsilon$ (up to $O(\epsilon)$, in fact)
that has caused a
finite shift proportional to $\pi^2$ in the NNLO coefficient.
The operators $\{O_l,O_p\}$
do not mix with each other at the $\alpha_s$ order.

The results~(\ref{eq:cp}), (\ref{eq:cl}) can be re-written at $N=3$, $n_l=4$ as
\begin{align*}
&C_p^{(2)} = - 25.333 \beta_0 - 100.055 = - 211.111 - 100.055 = - 311.166\,,\\
&C_l^{(2)} = - 43.906 \beta_0 + 190.323 = - 365.882 + 190.323 = - 175.560\,.
\end{align*}
So, the naive non-abelianization~\cite{BG:95} works moderately well:
the $\beta_0$ terms are about 2 times larger than those without $\beta_0$.

The results for the matching coefficients
correspond to the normalization point $m_b\equiv m_b^{pole}$,
$\mu = m_b^{pole}$.
One sees that the NNLO corrections are rather large.
Their calculation is crucial for estimating the reliability of the
results at NLO in PT.
We discuss some physical implications of the obtained results later.

\section{Description of computation}
\label{Sect:comp}

QCD operators can be expanded in $1/m$ in terms of HQET operators:
\begin{equation}
Q(\mu) = C(\mu) O(\mu)
+ \frac{1}{m} B(\mu) P(\mu)
+ \mathcal{O}\left(\frac{1}{m^2}\right)\,,
\label{calc:ope}
\end{equation}
where $Q(\mu)$ is the column of renormalized QCD operators,
$O(\mu)$ is the column of renormalized HQET operators,
and $C(\mu)$ is the matrix of matching coefficients.
For example, the leading matching coefficients $C$ for heavy--light quark currents
were calculated in~\cite{EH:90,BG:95,Grozin:1998kf,Bekavac:2009zc}.

Here we follow the same procedure but for the 4-quark operators.
First we calculate the bare matching coefficients
\begin{equation}
Q_0 = C_0 O_0\,,\quad
Q_0 = Z(\mu) Q(\mu)\,,\quad
O_0 = \tilde{Z}(\mu) O(\mu)\,,\quad
C(\mu) = Z^{-1}(\mu) C_0 \tilde{Z}(\mu)\,,
\label{calc:bare}
\end{equation}
where $Z(\mu)$ and $\tilde{Z}(\mu)$ are the matrices of renormalization constants in QCD and HQET
(we omit $1/m$ corrections).
This is done by equating the on-shell matrix element of $Q_0$
\begin{equation}
{<}\bar{b}d|Q_0|b\bar{d}{>} = Z_Q^{\text{os}} Z_q^{\text{os}} \Gamma_0\,,
\label{calc:qcd}
\end{equation}
where $Z_Q^{\text{os}}$ is the on-shell heavy-quark field renormalization constant,
$Z_q^{\text{os}}$ is the on-shell light-quark renormalization constant,
and $\Gamma_0$ is the vertex function of $Q_0$,
to $C_0$ times the corresponding on-shell matrix element of $O_0$.
If all light quarks (including the charmed quark $c$)
are considered massless,
all loop corrections to the HQET quantities $\tilde{Z}_Q^{\text{os}}$, $\tilde{Z}_q^{\text{os}}$, $\tilde{\Gamma}_0$
vanish in dimensional regularization because they contain no scale.

The 2-loop results for $Z_Q^{\text{os}}$~\cite{Broadhurst:1991fy}
and $Z_q^{\text{os}}$~\cite{Grozin:1998kf} are known.
We calculate $\Gamma_0$ up to 2 loops using the \textsc{reduce}
package
\textsc{recursor}~\cite{Broadhurst:1991fi},
similarly to~\cite{BG:95}.
The basis of antisymmetrized products of $\gamma_\bot$
allows us to project onto individual HQET operators easily.
The calculation is done in an arbitrary covariant gauge;
we check that the on-shell matrix element~(\ref{calc:qcd}) is gauge invariant
($Z_Q^{\text{os}}$ and $Z_q^{\text{os}}$ are gauge invariant up to 2 loops).
If we keep only the (gauge-invariant) subset of diagrams
in which only $b$ quark and the light
quark belonging to the same color-singlet current interact,
we successfully reproduce the 2-loop matching coefficients~\cite{BG:95,Grozin:1998kf}
of the heavy--light currents with the Dirac structures $1$, $\rlap/v$, $\gamma_\bot$.

The quantities $\Gamma_0$, $Z_Q^{\text{os}}$, $Z_q^{\text{os}}$ are calculated via the $n_f$-flavor bare coupling $g_0^{(n_f)}$;
we re-express them via the renormalized $\alpha_s^{(n_f)}(\mu)$.
The renormalization constant matrix $Z(\mu)$ is also expressed via $\alpha_s^{(n_f)}(\mu)$.
However, the renormalization constant matrix $\tilde{Z}(\mu)$ is expressed via $\alpha_s^{(n_l)}(\mu)$, $n_l=n_f-1$.
We have to use the decoupling relation between $\alpha_s^{(n_f)}(\mu)$ and $\alpha_s^{(n_l)}(\mu)$,
see, e.\,g., an introductory review~\cite{Grozin:2012ec}.
It is most convenient to perform matching at $\mu=m$, the
on-shell mass of the $b$ quark.
Then
\begin{equation}
\alpha_s^{(n_f)}(m) = \alpha_s^{(n_l)}(m) \left[ 1
+ T_F \frac{\pi^2}{9} \varepsilon \frac{\alpha_s^{(n_l)}(m)}{4\pi} \right]\,
\label{calc:dec}
\end{equation}
with $T_F=T(R)=1/2$ for fermions in fundamental representation.
We need this $\mathcal{O}(\varepsilon)$ term because
the 1-loop QCD matrix element contains poles in $1/\varepsilon$.

The computation has been done in leading logs 
in~\cite{PW:88,SV:88} where LO anomalous dimension for $O_l$ has been
found.
It happens to be equal to that of the simple product of 
two heavy light currents.
In higher orders this factorization does not necessarily persist. 
The standard result for the coefficients $C_{l,p}$
at NLO has been obtained in~\cite{FHH:91,CFG:96,B:96}.

In our paper~\cite{Grozin:2017uto}
we have calculated the NNLO contribution of
the leading order in $n_l$ only.
This allows one to estimate the full results
for the 2-loop matching coefficients using the approximate method of naive
non-abelianization~\cite{BG:95}.
Presently available
techniques allow for the analytical calculation of any number
of fermionic bubbles~(e.\,g.\ Chapher~8
in~\cite{G:04}, also available as~\cite{Grozin:2003gf})
that can be converted
into the $\beta_0$-dominance estimates of the matching coefficients
at any order of perturbative expansion. While the estimate is
technically feasible,
the quantitative validity of the approximation for a phenomenological
analysis
is not immediately clear
(e.\, g. see some discussion in~\cite{Krasnikov:1996jq}).

General description of matching calculations given above 
is well known. In our case of matching four-quark operators of QCD
onto four-quark operators in HQET there is a subtlety of using
dimensional regularization caused by the presence
of spurious operators that formally vanish in four-dimensional space.
Therefore one organizes the basis of operators in both QCD and HQET
in physical and evanescent sectors.
In QCD the operators are symbolically
$Q=\{Q,E\}$ (do not confuse the concrete operator $Q$ and general
notation for the whole set in QCD). In HQET the operators
are  $O=\{O_l,O_p,e_i\}$.
The matching is by necessity a relation between whole basis sets of the
operators in QCD and in HQET and it reads in general matrix form
\begin{eqnarray}  
Q=C\, O\, .
  \label{eq:matchgen}
\end{eqnarray}  
Here $C O$ denotes a matrix multiplication of the matrix of
matching coefficients $C$ and the string (one dimensional matrix) of
operators of the basis.

The renormalization pattern of
the operators in eq.~(\ref{eq:matchgen}) (including evanescent ones)
  is 
\begin{eqnarray}  
  Q=Z^{-1} Q^B,\quad \quad O={\tilde Z}^{-1} O^B
  \label{eq:renormgen}
\end{eqnarray}  
with $Z,\tilde Z$ being the renormalization matrices and $Q^B,O^B$ are bare
images of renormalized operators in QCD and HQET.
One obtains for the matching in eq.~(\ref{eq:matchgen})
\begin{eqnarray}  
Q^B=ZC{\tilde Z}^{-1}O^B\, .
\end{eqnarray}  
The matching coefficients can be found by
taking the matrix elements on shell in HQET
from the both sides of this relation.
And for HQET one has
\begin{eqnarray}  
\langle O^B\rangle = \text{tree level values only}
\label{eq:matrHQETvan}
\end{eqnarray}  
because all loops are scaleless and vanish in dimensional regularization.
This is independent of whether the operator is physical one or
evanescent one.
Let $O|_j$ be an operation of projecting on a particular state $P_j$
(operator)
and one chooses a complete system of $P_j$  for the HQET basis such
that 
$O_i^B|_j=\delta_{ij}$.
For the basis of four quark operators this operation has a simple
realisation:
one takes traces in both Dirac strings. Since the bare operator of the
basis has a structure of a direct product $O_n^B=A_n\otimes B_n$,
symbolically,
\begin{eqnarray} 
  O_i^B|_j={\rm tr}(\gamma_{\perp}^{(j)}A_i)
  {\rm tr}(B_i\gamma_{\perp}^{(j)})\sim
  \delta_{ij}\, .
\end{eqnarray}  
The matching coefficients $C$ become 
\begin{eqnarray}  
C_{mn}=(Z^{-1})_{np}(Q_p^B|_j){\tilde Z}_{jm} \, .
\end{eqnarray}  
This is a working formula.
The quantity $(Q_p^B|_j)$ is basically a bare matching coefficient of
the bare operator $Q_p^B$
from the QCD basis to the bare operator $O_j^B$ in an
HQET basis. The bare matching coefficient depends on
one scale $m_b$ and is represented by loop integrals on
shell. At NNLO they are two loop integrals.

At one loop level the renormalization matrix in QCD reads
\begin{equation}
Z = 1 + \frac{\alpha_s}{4\pi\varepsilon}
\left(\begin{array}{cc}
z_{QQ} & z_{QE} \\
\varepsilon z_{EQ} & z_{EE}
\end{array}\right)\,,
\end{equation}
where $z_{EQ}$ is obtained from the requirement
that matrix elements of the renormalized evanescent operator $E(\mu)$
vanish,
$z_{EQ}=3(1-1/N)(3N+14-22/N)$.
Here $z_{QQ}=-3(1-1/N)$ is related to
an anomalous dimension of the physical
operator $Q$. It is independent of details of the basis.
The quantity $z_{QE}=-T_F$ describes the admixture of the evanescent
operator $E$ to the physical one. The quantity $z_{EE}$ is anomalous
dimension of evanescent operators. Note that there are more than
one evanescent operator in the basis, however for our computation we
need only one independent combination of those, and  $z_{EE}$ is
irrelevant altogether.
At the two-loop level we need only one additional entry
to the renormalization constant $Z$ which is $z_{QQ}^{(2)}$
\begin{equation}
Z = 1 + \frac{\alpha_s}{4\pi\varepsilon}
\left(\begin{array}{cc}
z_{QQ} & z_{QE} \\
\varepsilon z_{EQ} & z_{EE}
      \end{array}\right)+\left(\frac{\alpha_s}{4\pi\varepsilon}\right)^2
\left(\begin{array}{cc}
z_{QQ}^{(2)} & * \\
*& *
      \end{array}\right)
   \,,
\end{equation}
The value of  $z_{QQ}^{(2)}$
is reconstructed from a two-loop anomalous dimension of the
physical operator $Q$.
The anomalous dimension of the operator $Q(\mu)$ is
\begin{eqnarray}  
  -\frac{1}{2}\gamma_Q=z_{QQ} \frac{\alpha_s}{4\pi}
  +\left(\frac{\alpha_s}{4\pi}\right)^2\{
  \frac{1}{\ep}
  \left(
  \beta_0 z_{QQ} +  2 z_{QQ}^{(2)} - z_{QQ}^2 \right)- z_{QE}z_{EQ}\}
\end{eqnarray} 
with
\begin{equation}
  \beta_0=\frac{11}{3}N-\frac{2}{3}n_f
\end{equation}
and $n_f=n_l+1$ is the number of flavors.
For the perturbative expansion of the anomalous dimension of the form 
\begin{equation}
  \gamma(\alpha_s)
  = \frac{\alpha_s}{4\pi}\gamma^0
  +\left( \frac{\alpha_s}{4\pi}\right)^2\gamma^1
\end{equation}
the coefficient $\gamma^1$ in the basis described
reads~\cite{Buras:1989xd}
\begin{equation}
  \label{eq:gam1QCD}
  \gamma^1=\frac{N - 1}{2N}\left(-21 + \frac{57}{N} - \frac{19}{3}N
    +  \frac{4}{3}n_f\right)\, .
\end{equation}  

In QCD one needs only one evanescent operator for the whole
computation of matching in two-loops,
just that one which admixes to $Q$ at NLO even if it can be composed of
several basis operators $Q_n$.  

In HQET the renormalization matrix for relevant operators looks
similar
to that one in QCD
\begin{equation}
{\tilde Z} = 1 + \frac{\alpha_s}{4\pi\varepsilon}
\left(\begin{array}{cc}
z_{OO} & z_{Oe} \\
\varepsilon z_{eO} & z_{ee}
      \end{array}\right)+\left(\frac{\alpha_s}{4\pi\varepsilon}\right)^2
\left(\begin{array}{cc}
z_{OO}^{(2)} & * \\
*& *
      \end{array}\right)
   \,,
\end{equation}
but now the quantity $z_{OO}$ is a $2\times 2$ matrix in the subspace
of physical operators $\{O_l,O_p\}$.

In HQET two evanescent operators are necessary for renormalization
of physical operators,
one for $O_l$ and one
for $O_p$. For computing the matching coefficients the
whole set of evanescent operators is relevant or at least the one what
$Q$ can match onto at NLO
(in fact, there is one operator that is multiplied by the poles
of the matching coefficient and the other with only
finite parts).
It shows that, indeed, in general the whole operator
basis should be matched onto.

As for bare coefficients we need two-loop values for $C(Q\to O_l,O_p)$ and
one-loop values for $C(Q\to j)$ for any operator $O_l,O_p,e_i$ and
$C(E\to l,p)$. One more ingredient is $Z_Q$ on shell at NNLO 
from~\cite{Broadhurst:1991fy}.
The calculation contains rather delicate cancellations of infinities
(poles in $\ep$)
and fixing the finite parts according to sophisticated conventions. To
give just 
an example, the renormalization matrix $Z$ is expanded, by convention,
over the renormalized coupling $\alpha_s^{(n_l+1)}(\mu)$ while
the renormalization matrix $\tilde Z$ is expanded, by convention,
over the renormalized coupling $\alpha_s^{(n_l)}(\mu)$.
In the course of computation of
matching coefficients the poles in $\ep$ cancel.

We have computed the necessary quantities.
The quantity $z_{QQ}^{(2)}$ can be extracted from earlier calculations
through anomalous dimensions of $Q$ at two loops.
The matrix  $z_{OO}^{(2)}$ is related to the anomalous dimension
matrix of the
physical pair $(l,p)$. One of the entry has been considered
in~\cite{G:93}
where the result for
the anomalous dimension of $O_l$ has been given.
The whole basis of
operators was not explicitly specified in~\cite{G:93}.
We do not compute the corresponding anomalous dimension independently.
It can be extracted from our calculations though.
In fact, one can extract only the difference
$\gamma^1-{\tilde\gamma^1}$
that reads
\[
  {\tilde\gamma}^1-\gamma^1=\left(1-\frac{1}{N}\right)
  \left(n_l\left(3 + \frac{5}{3} N\right)
    +\pi^2\frac{2}{3N} (N-1) (2 + 2 N + N^2)\right.
\]
\[
\left. -\frac{177 + 161 N - 3 N^2 + 83 N^3}{12 N}\right)\, .
\]
Assuming the value for $\gamma_1$ from
eq.~(\ref{eq:gam1QCD}) with $n_f=n_l+1$
we have extracted the anomalous dimension ${\tilde\gamma}_1$ of
the HQET operator $O_l$
  \[
{\tilde\gamma}^{(1)}=\left(1-\frac{1}{N}\right)
\left( - \frac{-165+279 N+35 N^2+83 N^3}{12 N}
+  n_l\frac{11+5 N}{3}\right.
\]
\[
+\left. \pi^2\frac{2 (N-1) \left(2+2 N+N^2\right)}{3 N}\right)\, .
\]

The coefficient $n_l$ agrees with our paper~\cite{Grozin:2017uto}.
The whole expression disagree with the results quoted in
ref.~\cite{G:93}.
Numerically we find 
\[
  {\tilde\gamma}^{(1)}=\frac{2}{27} (-807 + 78\, n_l - 68\, \pi^2)
  = -86.3802\ldots
\]
while the result of ref.~\cite{G:93} reads
\[
  {\tilde\gamma}^{(1)}|_{\rm G}=\frac{2}{27} (-1212 + 96 n_l - 26\pi^2)
  =-80.3415\ldots
\]
for $n_l=4$ and QCD with $N=3$.
The irony of life is that the two
quantities ${\tilde\gamma}^{(1)}$ and ${\tilde\gamma}^{(1)}|_{\rm G}$
are rather close numerically though the analytical expressions for
them are quite different.

As for the calculation at
one loop level the whole program of renormalization of bases
of four quark operators in
QCD and HQET and matching onto one another has been explicitly
realized in a closed form. The corresponding
formulas are given in the Appendix.

\section{Implications for phenomenology of the mixing}
\label{Sect:pheno}
The splitting between the mass eigenstates of
$B^0$-$\bar{B}^0$
system is
determined, in the SM, by the non-diagonal matrix element of the  effective
Hamiltonian~(\ref{eq:effH})
\begin{equation}
  \Delta\, m= \langle B^0|{\cal H}| {\bar B}^0\rangle
  =C(m_W, m_t,\mu, \alpha_s(\mu)) \langle B^0|Q(\mu)|  {\bar B}^0  \rangle
\label{eq:delatM}
\end{equation}
or, more specifically, if the concrete calculations are made within
dimensional regularization, by the expression~(\ref{eq:EffHevan}).
After the coefficient function have been determined in  QCD
perturbation theory the
task is the computation of hadronic matrix
elements of four-quark operators, ${Q,E}$. While for evanescent
operators $E$ this task can be trivially accomplished since by
construction $\langle B^0|E(\mu)|  {\bar B}^0  \rangle = 0$,
an accurate
determination of numerical values for physical operators is a
genuine
QCD-bound-states problem. The nonperturbative techniques required for
the computation can be QCD sum rules
and direct lattice simulation.
The sum rules results for the matrix element
$\langle B^0|Q(m_b)|  {\bar  B}^0\rangle $ have been presented
earlier~\cite{Narison:1994zt,Korner:2003zk,Ovchinnikov:1988zw,MPP:11}
within different approximation schemes 
for the Green functions used in the OPE.
The obtained  precision has been rather limited by modern standards
though.
Till recently the lattice analysis with fully relativistic
heavy quarks was impossible due to
insufficient computational power,
therefore the 
matching to HQET~(\ref{eq:match}) has been a necessity.
The NLO results for matching
coefficients to both HQET and to the lattice representation for the
operators have been
obtained more than a quarter of a century ago.

The technical breakthrough with  computation of
three loop HQET integrals in
ref~\cite{GL:09}
allowed for a NLO analysis of
the mixing matrix element using sum rules in HQET. 
In our recent paper~\cite{Grozin:2016uqy} we
have computed the bag parameter $B_d$ for $B^0_d-{\bar B}^0_d$
mixing at the next-to-leading order of perturbative expansion for matching
coefficient and for the Green functions entering the sum rules analysis. 
To evaluate the matrix element of the mixing we use 
a vertex (three-point) correlation function~\cite{Chetyrkin:1985vj}.
The analysis uses the  splitting of the whole Green function
necessary for the
calculation within OPE and for the sum rule
approach~\cite{Chetyrkin:1985vj,Ovchinnikov:1988zw,MPP:11}
into factorizable and
non-factorizable parts.
It happens that the non-factorizable part starts only
at NLO of perturbation
theory and turns out to be small.
These features allow for getting a numerical result of high
precision
for the bag parameter. The techniques have been also used for other
four quark operators in~\cite{Kirk:2017juj}\footnote{
  Note a minor misrepresentation of the results 
  in~\cite{Kirk:2017juj}: the right-hand side of eq.~(3.17)
  should include the factor $N C_F/4$ for general values of $N$.
  The results are correct in QCD for $N=3$ since $N C_F/4=1$.
  A.Lenz has confirmed
  this finding in his private communication to us.}.
The computation of
ref.~\cite{Kirk:2017juj} is pinning down the important uncertainty for
lifetime differences of the heavy mesons with both $b$ and $c$ quarks.
It is rather a complete analysis but it is limited to only
NLO of the perturbation theory for HQET Green functions.
Future experimental data may require even more accurate theoretical
predictions.
For obtaining still better theoretical accuracy,
the NNLO perturbative corrections to matching
coefficients can be useful.
The first step in
this direction has been done in our recent paper~\cite{Grozin:2017uto}
where
the NNLO corrections proportional to
the $n_l$ factor
have been computed that allowed us to perform  an
approximate evaluation of the coefficients within the naive
non-abelianization ($\beta_0$ dominance) 
approach~\cite{BG:95}. In the present paper we have computed the full
NNLO
results for matching coefficients. They read numerically 
\begin{eqnarray}
  C_l(m_b)&=&1-12 a_s- 175.6a_s^2\, ,\\
C_p(m_b)&=&-8 a_s- 311.2 a_s^2\, ,
  \end{eqnarray}
where $a_s=\alpha_s^{(4)}(m_b)/(4\pi)$.
Our calculation of matching coefficients is an important step in the
program of NNLO description of mixing within analytical methods
of computation.
From our results we see that NLO
approximation for matching coefficient may not be reliable
since the NNLO corrections are large.
On the other hand, we also see that the values of corrections to the
coefficients by
themselves do not lead to immediate physical conclusions.
One has to add corrections to corresponding Green functions
which determine the OPE for sum rule
analysis.
Thus, the NLO correction to $C_l$ requires NLO correction to the
correlator 
\begin{equation}
K = \int d^d x_1\,d^d x_2\,e^{i p_1 x_1 - i p_2 x_2} 
\langle 0 |T \tilde{\jmath}_2(x_2) 
O_l(0) \tilde{\jmath}_1(x_1) | 0 \rangle 
\label{SR:K}
\end{equation}
of the operator $O_l$ for consistent NLO analysis (see
ref.~\cite{Grozin:2017uto}
for more details).
In case of the $O_p$ operator the corresponding correlator of the $O_p$
operator can be taken at tree-level approximation
for obtainig the NLO result for the QCD
matrix element.
With the NNLO correction included to $C_p$, one
needs the NLO correction to the correlators with insertion
of $O_p$; this
turns out to be feasible~\cite{GL:09}
and it is even available for some cases~\cite{Kirk:2017juj}. 
In case of the NNLO corrections to $C_l$ one has to compute the NNLO
corrections to the
correlator~(\ref{SR:K}) as well 
that leads to four-loop integrals, which are currently beyond known
technology.

Note that with at NNLO there will appear new evanescent operators
which we do not specify explicitly. 
However, one can
try to choose the basis of evanescent operators such that  $C_l^{(2)}$
becomes smaller or even vanishing (that will change the numerical value of
$C_p^{(2)}$ accordingly) and one can hope to obtain a chance to
avoid the necessity of the computation of NNLO
corrections to the correlator~(\ref{SR:K}).

The other possibility to get the most of our NNLO calculation of
matching coefficients even without
having the NNLO results for the correlator~(\ref{SR:K})
is to perform a direct comparison between physical 
quantities. While most observables in QCD contain non-perturbative
contributions, 
there are, in fact, some observables for which one can show that one may 
construct a perturbative relation between
the two observables. In such a case the 
perturbative expansion acquires an immediate physical meaning
and statements about 
the size of coefficients as well as on
the convergence of the perturbative series become 
meaningful. To this end, while for the individual
observables large coefficients may appear 
(depending on the definition of the operator matrix elements),
in a relation between 
these observables the large coefficients my cancel,
once the matrix elements are defined 
in the same way in both observables. 

As we pointed out above, the bag factor $B_B$ turns out to be small
and hence to a good 
approximation we expect a perturbative relation between
$\Delta M_d$ and the $B$-meson
decay constant $f_B$. In fact, it is known that the matching coefficients
of the axial current are also large~\cite{BG:95}, and we may
expect that in a direct comparison 
of the axial vector current with $\Delta M_d$ a well-behaved
perturbation theory results. 

The matching coefficient of the axial current to HQET interpolating
operator is~\cite{BG:95,Grozin:1998kf}
\begin{align}
  J ={}& C_{ J} \tilde{\jmath}\,,\quad J
         =v_\mu J^\mu_5=\bar q\slashed{v}\gamma_5b\, ,
         \quad \tilde{\jmath}=\bar q\gamma_5h_v\, ,
\nonumber\\
C_J(m_b) ={}& 1 - 2 C_F a_s
+ C_F \biggl[ C_F \left(4 I_0 - 8 \zeta_3 - \frac{5}{2} \pi^2 + \frac{255}{16} \right)\nonumber\\
&{}+ C_A \left( - 2 I_0 + 2 \zeta_3 + \frac{5}{6} \pi^2 - \frac{871}{48} \right)
\nonumber\\
&{} + T_F n_l \left( \frac{2}{3} \pi^2 + \frac{47}{12} \right)
+ T_F \left( - 4 \pi^2 + \frac{727}{18} \right) \biggr] a_s^2
\nonumber\\
={}& 1 - \frac{8}{3} a_s - 31.6 a_s^2
\label{eq:cJ}
\end{align}
with a rather sizable coefficient at NNLO.

The HQET matching~(\ref{eq:match}) has a great deal of
arbitraryness in distributing the contributions between coefficients
and operators. One concrete choice of fixing the 
$\overline{\rm MS}$-scheme for the definition of the operators
is a current standard.
If we call $Q(m_b)$ to be our physical quantity determining $\delta m$
(again up to the freedom of redefinition the matrix elements and the
coefficients in QCD but we set this aside now)
then the expansion in HQET is not unique in a variety of aspects:
choosing a physical pair of the operators,
changing renormalization scheme, etc.
Also clearly, the freedom of the definition of evanescent
operators 
is not only the choice of a physical pair but
deviation from minimality.
By adding a physical operator to an evanescent one
with a coefficient vanishing at $\epsilon=0$
we obtain different renormalized physical operators,
see~(\ref{PhysEva}) and therefore a different matching coefficient.
 
While physics, i.e. predictions for observables,
does not depend on this reshuffling of the operator bases, the
independence restores only 
after adding matrix elements computed within the same scheme up to
same accuracy. The magnitude of the correction to a particular
coefficient is not an invariant characteristics of PT expansion and one
should really collect Wilson coefficients and matrix elements
together which is difficult in QCD
since there is no any quantitative scheme for computing hadronic matrix
elements. Lattice simulation may be an exception, but then the
perturbative short
distance part of the matrix element cannot be easily identified.

The full NNLO analysis of the mixing seems to be feasible, and the
most intriguing part is the possibility to find a basis of evanescent
operators in which the NNLO correction to the matching coefficient of
the operator $O_l$ turns out to be small.
This assertion deserves to be validated in future work.

Note also that the two-loop anomalous dimensions of the operators
$\{O_l,O_p\}$
are not very important
quantitatively. The point is that the difference of scales is not
large and summation of logarithms of the scale differences is not
crucial numerically. Indeed, matching is done at $m_b$ with
$m_b^{pole}\sim 4.8~\text{GeV}$~(e.g.~\cite{Penin:1998kx,Hoang:2000yr})
while the Green
functions necessary for sum rules
analysis are computed at the scale of the order
of $w_0\sim 1~\text{GeV}$. The leading logs of the form
$(\alpha_s(w_0)\ln(w_0/m_b))^n$, with $n>0$ can be summed with the leading
anomalous dimensions of the operators $\{O_l,O_p\}$ which are known
while the subleading logs of the form 
$\alpha_s(w_0)(\alpha_s(w_0)\ln(w_0/m_b))^n$  with $n\ge 1$ 
are not large and can be retained in an expanded form.

Here we check that our results for matching coefficients allow for
a reasonable perturbative expansion of physical parameters.
Recall that the matrix element in QCD is represented by the expression
\begin{equation}
  \langle B^0|Q(\mu)|  {\bar B}^0  \rangle
  =2\left(1+\frac{1}{N}\right)B_q(\mu)f_B^2
\label{eq:factBq}
\end{equation}
and therefore $B_q(\mu)$ gives a relation between physical parameters
$\Delta m$ and $f_B^2$ (we factor out the redefinition freedom of the
operator $Q(\mu)$ in QCD
considering it fully under control in perturbation theory).
The perturbation theory series for the relation between 
$\Delta m$ and $f_B^2$ is 
unambiguous and can depend only on $N_c$ and $\alpha_s$ (with all
possible reservations that they are not completely PT quantities).
The realization of this idea is given in eq.~(2.18) of
ref.~\cite{Grozin:2017uto} in the form
\begin{equation}
  C_J^2 B_q(m_b) = C_l(m_b)B_l+ C_p\left(\frac{(-2N+1)}{2(N+1)}B_s
    +\frac{1}{4}B_l\right)
\label{eq:BqvsBlN}
\end{equation}
or
\begin{equation}
  B_q(m_b) = \{C_l(m_b)B_l+ C_p\left(-\frac{5}{8}B_s
    +\frac{1}{4}B_l\right)\}/C_J^2\, .
\label{eq:BqvsBl}
\end{equation}
Here $B_{l,s}$ are bag parameters for the HQET operators $O_{l,s}$
(see,
ref.~\cite{Grozin:2017uto} for more detail).
The perturbative corrections to the parameter $B_l$
has the form~\cite{Grozin:2016uqy}
\begin{eqnarray}
  B_l&\to &B_l\left(1 - a_s\frac{N - 1}{2N}(\frac{4}{3} \pi^2 - 5)
            - x_l^{(2)} a_s^2\right)\, ,\\
  \label{eq:deltaBl}
\end{eqnarray}
where $x_l^{(2)}$ represents an (yet unknown)
NNLO correction to $B_l$ from HQET
sum rules (compare to the parameter $X$ used in~\cite{Grozin:2016uqy}
for estimating higher order contributions).
We have extracted the correction to the parameter $B_s$ from
ref.~\cite{Kirk:2017juj} with the result
\begin{eqnarray}
B_s &\to & B_s\left(1 - a_s\frac{8}{15}\left(9 -
            \frac{2\pi^2}{3}\right)
            \right)\, .
           \label{eq:deltaBs}
\end{eqnarray}
Substituting our results for the coefficients
$C_l,C_p$ and the expression for $C_J$,
we
finally obtain the relation between physical observables in the form 
\begin{eqnarray}
  \Delta m=\text{const}
  \left(1 - 6.4 a_s - (4.9 +  x_l^{(2)})a_s^2\right)f_B^2\, .
  \label{eq:finalPT}
\end{eqnarray}
The quantity $x_l^{(2)}$ emerges from NNLO corrections to HQET sum
rules and it 
is expected to be in the range of NLO correction which is just of
order unity (compare assumptions on the parameter $X$ 
in~\cite{Grozin:2016uqy}, $|X|<20$).
The perturbative expansion in eq.~(\ref{eq:finalPT}) has reasonably
small coefficients as we had expected.
We see that all large coefficients in $C_l,C_p,C_J$ mutually cancel
each other 
and numbers of order unity survive in the final expression
rending the PT expansion to be
reliable.

The conjecture about the above pattern of perturbative expansion
is quite feasible and can be explicitly checked once
$x_l^{(2)}$ is determined. Note that the idea of re-expressing the
physical quantities through one another with a resulting PT factor
is rather old and has been widely used. It is often applied as well to 
observables which are not quite fully perturbatively related (see,
e.g.~\cite{Penin:1998wj,Beneke:1998sy}.

Note that the lattice results are based on NLO analysis for the
matching coefficients between continuum and lattice representations of
the operators. While that matching is different, our computation shows
that PT corrections at NNLO to the matching coefficients
can be important at this level of
precision.

If there is no particularly large NNLO contribution to the sum rule
determination
of $x_l^{(2)}$,
our numbers show that
the precision of the matrix element
at the level of few percents even in the presence
of NNLO matching can be obtained. The large corrections are mainly
hidden in $C_J^2$. In other words, the single out of $B$ parameter is
very efficient physically since it has a reasonable perturbative
expansion {\it a
posteriori} and a bunch of perturbative
correction to the matrix element simply reproduces the
correct value of $f_B$.

Note in passing that with the NNLO accuracy of
the leading term one may need to account also for nonleading terms of
HQE~\cite{KM:92} while small
power corrections have been accounted
in~\cite{MPP:11}. 

In general, up-to-date lattice results turn out to be more precise
than sum rule estimates for classical quantities like decay
constants~\cite{Gelhausen:2013wia,Rosner:2015wva}.
Only due to a special structure of the
observables occurring in the $B^0-{\bar B}^0$
mixing
one can still use the sum rules to obtain predictions that are
still 
competitive with lattice computations.

Extension of our results to the case of $B^0_s {\bar B}^0_s $ has been
discussed in~\cite{Grozin:2017uto}. As the mass $m_s$ is not large
it can be taken into account in expansion in $m_s/w_0$ (an interesting
example is given in~\cite{Ovchinnikov:1985bs}).
We have shown that the strange
quark mass appears in non-factorizable quantities only at NLO level
while the leading order contributions are hidden in factorizable
parameters like $f_{B_s}$~\cite{Gelhausen:2013wia}.
Numerically the parameter $(m_s/w_0)\alpha_s(w_0)$~\cite{Korner:2000wd}
corresponds to NNLO
level 
but achieving this accuracy
requires only one loop calculations. With formulas given in the
Appendix it is rather a straightforward computation which we are going to
present in future publications.

\section{Summary}
\label{Sect:Summary}
We have calculated NNLO corrections to matching coefficients necessary
for the analysis of mixing in $B^0 - \bar B^0$ system, i.e. for
the calculation of bag parameters in sum rules at
three-loop level in HQET~\cite{Grozin:2016uqy,Kirk:2017juj}. 
The NNLO corrections happen to be large, however, to rather large
extent they cancel the large NNLO corrections in the matching
coefficients for the axial current that determines the $B$-meson
leptonic decay constant, $f_B$. The relation between
experimentally measured quantities $\Delta m$ and $f_B^2$
turns out to be rather well behaved as a perturbative series up to
NNLO.
This observation gives a strong ground to our estimate of
the uncertainties for the QCD bag parameter $B_q$ along the lines of
ref.~\cite{Grozin:2016uqy,Grozin:2017uto}.

We also discuss possible ways of getting invariant physical predictions
from our results independent of
the introduction of evanescent operators in the physical sector
due to renormalization.
We have constructed and present
the completely analytical framework for analyzing
the renormalization and matching of four quark operators at one loop
level.

\section{Acknowledgment}

A.G. is grateful to Siegen University for hospitality;
his work has been partially supported by the Russian Ministry of 
Education and Science.
This work is supported by the DFG Research Unit FOR 1873
"Quark Flavour Physics and Effective Theories".

\begin{appendices}

\section{Renormalization of four-quark operators at 1 loop:
  QCD}
Let us consider a bare operator
\begin{equation}
O_0 = \frac{1}{2} \bigl(\bar{d}_{L0i} \Gamma b_0^i\bigr) \bigl(\bar{d}_{L0j} \Gamma b_0^j\bigr)\,,
\label{QCD:O}
\end{equation}
where $\Gamma$ is an arbitrary Dirac matrix.
In the Born approximation its matrix element is
\begin{equation}
\begin{split}
{<}\bar{b}d|O_0|b\bar{d}{>}
&{} = \raisebox{-\figh}{\begin{picture}(22.6,22.6)
\put(11.3,11.3){\makebox(0,0){\includegraphics{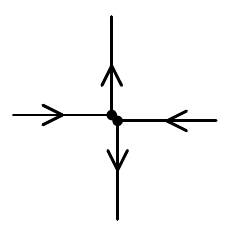}}}
\put(2.3,13.6){\makebox(0,0){$u_b$}}
\put(20.3,9){\makebox(0,0){$v_b$}}
\put(9,20.3){\makebox(0,0){$\bar{v}_d$}}
\put(13.6,2.3){\makebox(0,0){$\bar{u}_d$}}
\end{picture}}
- \raisebox{-\figh}{\begin{picture}(22.6,22.6)
\put(11.3,11.3){\makebox(0,0){\includegraphics{b0.pdf}}}
\put(2.3,13.6){\makebox(0,0){$u_b$}}
\put(20.3,9){\makebox(0,0){$v_b$}}
\put(9,20.3){\makebox(0,0){$\bar{u}_d$}}
\put(13.6,2.3){\makebox(0,0){$\bar{v}_d$}}
\end{picture}}\\
&{} = \bigl(\bar{v}_{di} \Gamma u_b^i\bigr) \bigl(\bar{u}_{dj} \Gamma v_b^j\bigr)
- \bigl(\bar{u}_{dj} \Gamma u_b^i\bigr) \bigl(\bar{v}_{di} \Gamma v_b^j\bigr)\,.
\end{split}
\label{QCD:0}
\end{equation}
For brevity we'll write it as $M_0 = T_1 \Gamma \otimes \Gamma$ where
\begin{equation}
\begin{split}
&\Gamma_1 \otimes \Gamma_2 \equiv
\bigl(\bar{v}_{d i_1} \Gamma_1 u_b^{j_1}\bigr) \bigl(\bar{u}_{d i_2} \Gamma_2 v_b^{j_2}\bigr)
- \bigl(\bar{u}_{d i_1} \Gamma_1 u_b^{j_1}\bigr) \bigl(\bar{v}_{d i_2} \Gamma_2 v_b^{j_2}\bigr)\,,\\
&T_1 \equiv \delta^{i_1}_{j_1} \delta^{i_2}_{j_2}\,,\qquad
T_2 \equiv \delta^{i_1}_{j_2} \delta^{i_2}_{j_1}\,.
\end{split}
\label{QCD:Brevity}
\end{equation}

The 1-loop matrix element is
\begin{equation}
\begin{split}
&{<}\bar{b}d|O_0|b\bar{d}{>} = Z_q^2 \Biggl[
\raisebox{-\figh}{\begin{picture}(22.6,22.6)
\put(11.3,11.3){\makebox(0,0){\includegraphics{b0.pdf}}}
\end{picture}}
+ \raisebox{-\figh}{\includegraphics{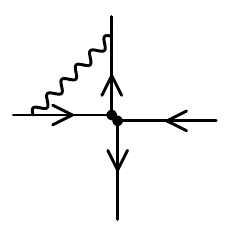}}
+ \raisebox{-\figh}{\includegraphics{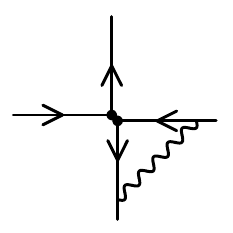}}\\
&{} + \raisebox{-\figh}{\includegraphics{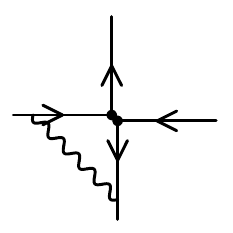}}
+ \raisebox{-\figh}{\includegraphics{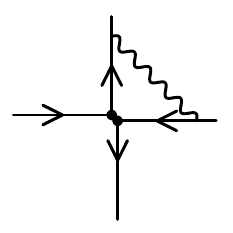}}
+ \raisebox{-\figh}{\includegraphics{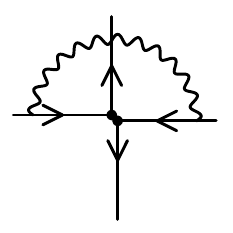}}
+ \raisebox{-\figh}{\includegraphics{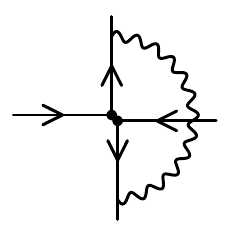}}
\Biggr]\\
&{} = Z_q^2 \bigl[M_0 + M_1 + M_2 + M_3 + M_4 + M_5 + M_6\bigr]\,.
\end{split}
\label{QCD:me}
\end{equation}
We are interested only in the UV $1/\varepsilon$ divergent terms.
Therefore we may treat all quarks as massless and set all external momenta to 0.
Then we need some IR regulator, say,
replacing all massless denominators by the ones with some non-zero mass,
or a hard IR cutoff in euclidean momentum space.
Such a regulator will be implied, not written explicitly.
The $\overline{\text{MS}}$ quark field renormalization constant is
\begin{equation}
Z_q = 1 - C_F \frac{\alpha_s}{4\pi\varepsilon} (1 - \xi)\,,
\label{QCD:Zq}
\end{equation}
where $\xi$ is the gauge fixing parameter.

Averaging over directions of the loop momentum $k$ in the integrands, we easily obtain
\begin{align*}
&M_1 = C_F T_1 \frac{\alpha_s}{4\pi\varepsilon}
\left[ \frac{1}{d} \gamma^\mu \gamma^\nu \Gamma \gamma_\nu \gamma_\mu \otimes \Gamma
- \xi \Gamma \otimes \Gamma \right]\,,\\
&M_2 = C_F T_1 \frac{\alpha_s}{4\pi\varepsilon}
\left[ \frac{1}{d} \Gamma \otimes \gamma^\mu \gamma^\nu \Gamma \gamma_\nu \gamma_\mu
- \xi \Gamma \otimes \Gamma \right]\,,\\
&M_3 = T_F \left(T_2 - \frac{T_1}{N}\right) \frac{\alpha_s}{4\pi\varepsilon}
\left[ \frac{1}{d} \Gamma \gamma^\nu \gamma^\mu \otimes \gamma_\mu \gamma_\nu \Gamma
- \xi \Gamma \otimes \Gamma \right]\,,\\
&M_4 = T_F \left(T_2 - \frac{T_1}{N}\right) \frac{\alpha_s}{4\pi\varepsilon}
\left[ \frac{1}{d} \gamma^\mu \gamma^\nu \Gamma \otimes \Gamma \gamma_\nu \gamma_\mu
- \xi \Gamma \otimes \Gamma \right]\,,\\
&M_5 = - T_F \left(T_2 - \frac{T_1}{N}\right) \frac{\alpha_s}{4\pi\varepsilon}
\left[ \frac{1}{d} \Gamma \gamma^\nu \gamma^\mu \otimes \Gamma \gamma_\nu \gamma_\mu
- \xi \Gamma \otimes \Gamma \right]\,,\\
&M_6 = - T_F \left(T_2 - \frac{T_1}{N}\right) \frac{\alpha_s}{4\pi\varepsilon}
\left[ \frac{1}{d} \gamma^\mu \gamma^\nu \Gamma \otimes \gamma_\mu \gamma_\nu \Gamma
- \xi \Gamma \otimes \Gamma \right]\,.
\end{align*}
The matrix element~(\ref{QCD:me}) is gauge invariant:
\begin{equation}
\begin{split}
&{<}\bar{b}d|O_0|b\bar{d}{>} = T_1 \Gamma \otimes \Gamma\\
&{} + C_F T_1 \frac{\alpha_s}{4\pi\varepsilon}
\Bigl[ \frac{1}{d} \gamma^\mu \gamma^\nu \Gamma \gamma_\nu \gamma_\mu \otimes \Gamma
+ \frac{1}{d} \Gamma \otimes \gamma^\mu \gamma^\nu \Gamma \gamma_\nu \gamma_\mu
- 2 \Gamma \otimes \Gamma \Bigr]\\
&{} + T_F \left(T_2 - \frac{T_1}{N}\right) \frac{\alpha_s}{4\pi\varepsilon} \frac{1}{d}
(\Gamma \gamma^\nu \gamma^\mu - \gamma_\mu \gamma_\nu \Gamma) \otimes
(\gamma_\mu \gamma_\nu \Gamma - \Gamma \gamma_\nu \gamma_\mu)\,,
\end{split}
\label{QCD:s}
\end{equation}
where only the UV $1/\varepsilon$ divergences are kept in the 1-loop terms.
For the operator
\begin{equation}
O'_0 = \frac{1}{2} \bigl(\bar{d}_{L0i} \Gamma b_0^j\bigr) \bigl(\bar{d}_{L0j} \Gamma b_0^i\bigr)
\label{QCD:Op}
\end{equation}
the only difference is that the color factors:
\begin{equation}
\begin{split}
&{<}\bar{b}d|O'_0|b\bar{d}{>} = T_2 \Gamma \otimes \Gamma\\
&{} + C_F T_2 \frac{\alpha_s}{4\pi\varepsilon}
\Bigl[ \frac{1}{d} \Gamma \gamma^\nu \gamma^\mu \otimes \gamma_\mu \gamma_\nu \Gamma
+ \frac{1}{d} \gamma^\mu \gamma^\nu \Gamma \otimes \Gamma \gamma_\nu \gamma_\mu
- 2 \Gamma \otimes \Gamma \Bigr]\\
&{} + T_F \left(T_1 - \frac{T_2}{N}\right) \frac{\alpha_s}{4\pi\varepsilon} \frac{1}{d}
\Bigr[ \gamma^\mu \gamma^\nu \Gamma \gamma_\nu \gamma_\mu \otimes \Gamma
+ \Gamma \otimes \gamma^\mu \gamma^\nu \Gamma \gamma_\nu \gamma_\mu \\
&\qquad{} - \Gamma \gamma^\nu \gamma^\mu \otimes \Gamma \gamma_\nu \gamma_\mu
- \gamma^\mu \gamma^\nu \Gamma \otimes \gamma_\mu \gamma_\nu \Gamma \Bigr]\,.
\end{split}
\label{QCD:ps}
\end{equation}

Now we specifically consider the operators
\begin{equation}
O_{n0} = \frac{1}{2} \bigl(\bar{d}_{L0i} \Gamma_n b_0^i\bigr) \bigl(\bar{d}_{L0j} \Gamma_n b_0^j\bigr)\,,\qquad
O'_{n0} = \frac{1}{2} \bigl(\bar{d}_{L0i} \Gamma_n b_0^j\bigr) \bigl(\bar{d}_{L0j} \Gamma_n b_0^i\bigr)\,,
\label{QCD:On}
\end{equation}
where
\begin{equation}
\Gamma_n = \gamma^{[\mu_1} \cdots \gamma^{\mu_n]}
\label{QCD:Gamman}
\end{equation}
is the antisymmetrized product of $n$ $\gamma$ matrices.
We have~\cite{Dugan:1990df}
\begin{equation}
\begin{split}
&\gamma^\mu \Gamma_n \gamma_\mu = (-1)^n (d - 2 n) \Gamma_n\,,\\
&\gamma^\mu \Gamma_n \otimes \gamma_\mu \Gamma_n =
\Gamma_{n+1} \otimes \Gamma_{n+1} + n (d-n+1) \Gamma_{n-1} \otimes \Gamma_{n-1}\,,\\
&\Gamma_n \gamma^\mu \otimes \gamma_\mu \Gamma_n =
(-1)^n \left[ \Gamma_{n+1} \otimes \Gamma_{n+1} - n (d-n+1) \Gamma_{n-1} \otimes \Gamma_{n-1} \right]\,,
\end{split}
\label{QCD:id}
\end{equation}
Using these relations twice, we can re-write the matrix
elements~(\ref{QCD:s}),
(\ref{QCD:ps}) as
\begin{align}
&{<}\bar{b}d|O_{n0}|b\bar{d}{>} = {<}O_n{>}
+ 2 C_F \frac{\alpha_s}{4\pi\varepsilon} \left[\frac{(d-2n)^2}{d} - 1\right] {<}O_n{>}
\label{QCD:mO}\\
&{} - T_F \frac{\alpha_s}{4\pi\varepsilon} \frac{4}{d}
\left[ {<}O'_{n+2}{>} - \frac{{<}O_{n+2}{>}}{N}
+ n (n-1) (d-n+1) (d-n+2) \left( {<}O'_{n-2}{>} - \frac{{<}O_{n-2}{>}}{N} \right)
\right]\,,
\nonumber\\
&{<}\bar{b}d|O'_{n0}|b\bar{d}{>} = {<}O'_n{>}
- C_F \frac{\alpha_s}{4\pi\varepsilon} \frac{2}{d}
\biggl[ {<}O'_{n+2}{>}
- 2 n (d-n) {<}O'_n{>}
\nonumber\\
&\qquad{} + n (n-1) (d-n+1) (d-n+2) {<}O'_{n-2}{>} \biggr]
\nonumber\\
&{} - T_F \frac{\alpha_s}{4\pi\varepsilon} \frac{2}{d}
\biggl[ {<}O_{n+2}{>} - \frac{{<}O'_{n+2}{>}}{N}
\nonumber\\
&\qquad{} + \bigl(6 n (d-n) - d (d-1)\bigr) \left( {<}O_n{>} - \frac{{<}O'_n{>}}{N} \right)
\nonumber\\
&\qquad{} + n (n-1) (d-n+1) (d-n+2) \left( {<}O_{n-2}{>} - \frac{{<}O'_{n-2}{>}}{N} \right)
\biggr]\,.
\label{QCD:mOp}
\end{align}
where ${<}O{>}$ in the right-hand side are the Born-level matrix elements.

\section{Renormalization of four-quark operators at 1 loop:
  HQET}

Let's consider the bare HQET operator
\begin{equation}
\tilde{O}_0 = \bigl(\bar{d}_{L0i} \Gamma h_{+0}^i\bigr) \bigl(\bar{d}_{L0j} \Gamma h_{-0}^j\bigr)\,,
\label{HQET:O}
\end{equation}
where $h_+$ annihilates a heavy quark and $h_-$ creates a heavy antiquark.
The 1-loop matrix element is
\begin{equation}
\begin{split}
&{<}\bar{b}d|\tilde{O}_0|b\bar{d}{>} = Z_q Z_h \Biggl[
\raisebox{-\figh}{\includegraphics{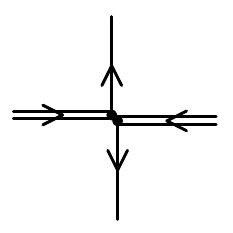}}
+ \raisebox{-\figh}{\includegraphics{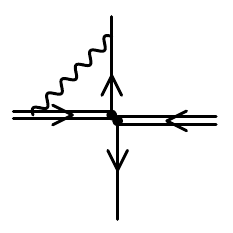}}
+ \raisebox{-\figh}{\includegraphics{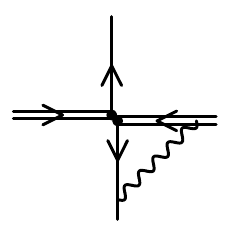}}\\
&{} + \raisebox{-\figh}{\includegraphics{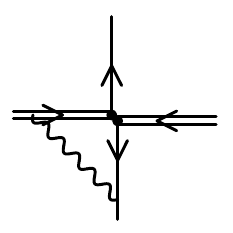}}
+ \raisebox{-\figh}{\includegraphics{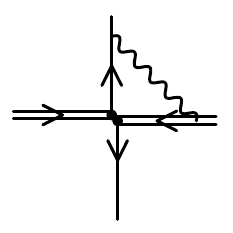}}
+ \raisebox{-\figh}{\includegraphics{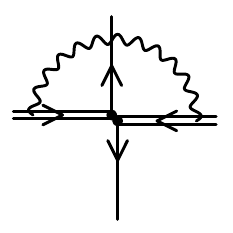}}
+ \raisebox{-\figh}{\includegraphics{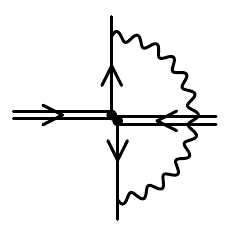}}
\Biggr]\\
&{} = Z_q Z_h \bigl[M_0 + M_1 + M_2 + M_3 + M_4 + M_5 + M_6\bigr]\,.
\end{split}
\label{HQET:me}
\end{equation}
We are interested only in the UV $1/\varepsilon$ divergent terms.
Therefore we may treat light quarks as massless, set their external momenta to 0
and external residual momenta of HQET (anti-) quarks to 0.
An IR cutoff is implied.
The $\overline{\text{MS}}$ HQET field renormalization constant is
\begin{equation}
Z_h = 1 + C_F \frac{\alpha_s}{4\pi\varepsilon} (2 + \xi)\,.
\label{HQET:Zh}
\end{equation}

Averaging the integrands over $k$ directions
(in particular, using $\overline{(k\cdot v)^{-2}}
= - (d-2) (k^2)^{-1}$~\cite{BG:95,GSS:06}),
we easily obtain
\begin{align*}
&M_1 = M_2 = C_F T_1 \frac{\alpha_s}{4\pi\varepsilon} (1-\xi) \Gamma \otimes \Gamma\,,\\
&M_3 = M_4 = T_F \left(T_2 - \frac{T_1}{N}\right) \frac{\alpha_s}{4\pi\varepsilon} (1-\xi) \Gamma \otimes \Gamma\,,\\
&M_5 = T_F \left(T_2 - \frac{T_1}{N}\right) \frac{\alpha_s}{4\pi\varepsilon} (d-2+\xi) \Gamma \otimes \Gamma\,,\\
&M_6 = - T_F \left(T_2 - \frac{T_1}{N}\right) \frac{\alpha_s}{4\pi\varepsilon}
\left[ \frac{1}{d} \gamma^\mu \gamma^\nu \Gamma \otimes \gamma_\mu \gamma_\nu \Gamma
- \xi \Gamma \otimes \Gamma \right]\,.
\end{align*}
The matrix element~(\ref{HQET:me}) is gauge invariant:
\begin{equation}
\begin{split}
&{<}\bar{b}d|\tilde{O}_0|b\bar{d}{>} = \biggl[ T_1
+ 3 C_F T_1 \frac{\alpha_s}{4\pi\varepsilon}
+ T_F \left(T_2 - \frac{T_1}{N}\right) \frac{\alpha_s}{4\pi\varepsilon} d \biggr]
\Gamma \otimes \Gamma\\
&{} - T_F \left(T_2 - \frac{T_1}{N}\right) \frac{\alpha_s}{4\pi\varepsilon} \frac{1}{d}
\gamma^\mu \gamma^\nu \Gamma \otimes \gamma_\mu \gamma_\nu \Gamma\,,
\end{split}
\label{HQET:s}
\end{equation}
where only the UV $1/\varepsilon$ divergences are kept in the 1-loop terms.
The matrix element of the operator
\begin{equation}
\tilde{O}'_0 = \bigl(\bar{d}_{L0i} \Gamma h_{+0}^j\bigr) \bigl(\bar{d}_{L0j} \Gamma h_{-0}^i\bigr)
\label{HQET:Op}
\end{equation}
differs from~(\ref{HQET:s}) only by the interchange $T_1\leftrightarrow T_2$.

Now we specifically consider the operators
\begin{equation}
\tilde{O}_{n0} = \bigl(\bar{d}_{L0i} \Gamma_{\bot n} h_{+0}^i\bigr) \bigl(\bar{d}_{L0j} \Gamma_{\bot n} h_{-0}^j\bigr)\,,\qquad
\tilde{O}'_{n0} = \bigl(\bar{d}_{L0i} \Gamma_{\bot n} h_{+0}^j\bigr) \bigl(\bar{d}_{L0j} \Gamma_{\bot n} h_{-0}^i\bigr)\,,
\label{HQET:On}
\end{equation}
where
\begin{equation}
\Gamma_{\bot n} = \gamma_\bot^{[\mu_1} \cdots \gamma_\bot^{\nu_n]}\,,\qquad
\gamma_\bot^\mu = \gamma^\mu - \rlap/v v^\mu\,.
\label{HQET:Gamman}
\end{equation}
Setting $\gamma^\mu = \rlap/v v^\mu + \gamma_\bot^\nu$ and using the $(d-1)$-dimensional versions of~(\ref{QCD:id}),
we can rewrite the matrix element~(\ref{HQET:s}) as
\begin{equation}
\begin{split}
&{<}\bar{b}d|\tilde{O}_{n0}|b\bar{d}{>} = {<}\tilde{O}_n{>}
+ 3 C_F \frac{\alpha_s}{4\pi\varepsilon} {<}\tilde{O}_n{>}\\
&{} - T_F \frac{\alpha_s}{4\pi\varepsilon} \frac{1}{d} \biggl[
{<}\tilde{O}'_{n+2}{>} - \frac{{<}\tilde{O}_{n+2}{>}}{N}
- 2 \biggl({<}\tilde{O}'_{n+1}{>} - \frac{{<}\tilde{O}_{n+1}{>}}{N}\biggr)\\
&\qquad{} + \bigl((d-1) (2n-d) - 2 n^2\bigr) \biggl({<}\tilde{O}'_n{>} - \frac{{<}\tilde{O}_n{>}}{N}\biggr)\\
&\qquad{} - 2 n (d-n) \biggl({<}\tilde{O}'_{n-1}{>} - \frac{{<}\tilde{O}_{n-1}{>}}{N}\biggr)\\
&\qquad{} + n (n-1) (d-n) (d-n+1) \biggl({<}\tilde{O}'_{n-2}{>} - \frac{{<}\tilde{O}_{n-2}{>}}{N}\biggr)
\biggr]\,.
\end{split}
\label{HQET:mO}
\end{equation}
The result for ${<}\bar{b}d|\tilde{O}'_{n0}|b\bar{d}{>}$ differs only by the interchange
of primed and non-primed operators in the right-hand side.

\section{Matching QCD on HQET at 1 loop}
\label{S:OS}
The 1-loop on-shell matrix element of the QCD operator~(\ref{QCD:O}) is
\begin{equation}
\begin{split}
&{<}\bar{b}d|O_0|b\bar{d}{>} = Z_Q^{\text{os}} \Biggl[
\raisebox{-\figh}{\includegraphics{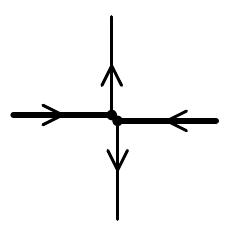}}
+ \raisebox{-\figh}{\includegraphics{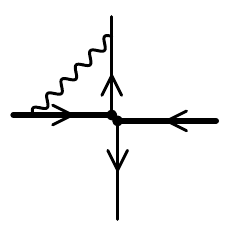}}
+ \raisebox{-\figh}{\includegraphics{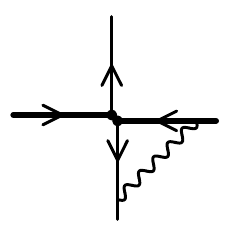}}\\
&{} + \raisebox{-\figh}{\includegraphics{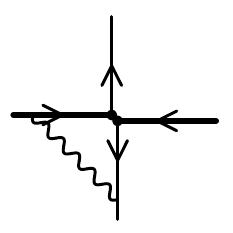}}
+ \raisebox{-\figh}{\includegraphics{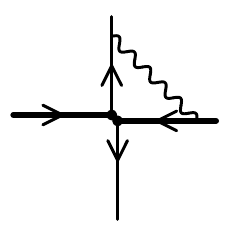}}
+ \raisebox{-\figh}{\includegraphics{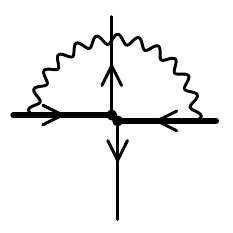}}
\Biggr]\\
&{} = Z_Q^{\text{os}} \bigl[M_0 + M_1 + M_2 + M_3 + M_4 + M_5\bigr]\,,
\end{split}
\label{OS:me}
\end{equation}
where
\begin{equation}
Z_Q^{\text{os}} = 1 - C_F \frac{g_0^2 m^{-2\varepsilon}}{(4\pi)^{d/2}} \Gamma(\varepsilon) \frac{d-1}{d-3}\,,
\label{OS:ZQ}
\end{equation}
$m$ is the on-shell mass, and $Z_q^{\text{os}}=1$ at this order.
The Born matrix element of the operator $O_{n0}$~(\ref{QCD:On}) is
$M_0 = T_1 \left(\Gamma_{\bot n} \otimes \Gamma_{\bot n} - n \Gamma_{\bot n-1} \otimes \Gamma_{\bot n-1}\right)$,
where we have used
\begin{equation}
\Gamma_n \otimes \Gamma_n = \Gamma_{\bot n} \otimes \Gamma_{\bot n} + n \Gamma_{\bot n-1} \rlap/v \otimes \Gamma_{\bot n-1} \rlap/v\,.
\label{OS:id}
\end{equation}

Each 1-loop diagram is gauge invariant separately:
\begin{align*}
&M_1 = M_2 = C_F T_1 \frac{g_0^2 m^{-2\varepsilon}}{(4\pi)^{d/2}} \Gamma(\varepsilon) \frac{1}{(d-2)(d-3)}
\Bigl[(d-n-1) (d-2n-2) {<}\tilde{O}_{n}{>}\\
&\qquad{} + n (n-1) (d-2n+2) {<}\tilde{O}_{n-1}{>}\Bigr]\,,\\
&M_3 = M_4 = T_F \left(T_2-\frac{T_1}{N}\right) \frac{g_0^2 m^{-2\varepsilon}}{(4\pi)^{d/2}} \Gamma(\varepsilon) \frac{1}{2(d-2)(d-3)}
\Bigl[- {<}\tilde{O}_{n+2}{>} + (d+n-2) {<}\tilde{O}_{n+1}{>}\\
&\qquad{} + n (3d-2n-4) {<}\tilde{O}_{n}{>} - n (d-n) (d+2n-4) {<}\tilde{O}_{n-1}{>}\\
&\qquad{} - n (n-1) (d-n+1) (2d-n-2) {<}\tilde{O}_{n-2}{>}\\
&\qquad{} + n (n-1) (n-2) (d-n+1) (d-n+2) {<}\tilde{O}_{n-3}{>}\Bigr]\,,\\
&M_5 = T_F \left(T_2-\frac{T_1}{N}\right) \frac{g_0^2 m^{-2\varepsilon}}{(4\pi)^{d/2}} \Gamma(\varepsilon) \frac{1}{2(d-2)}
\biggl[- {<}\tilde{O}_{n+2}{>} + (d+n-2) {<}\tilde{O}_{n+1}{>}\\
&\qquad{} + \frac{1}{d-3} \Bigl(-\bigl[2 (d-1) (d-2) + (d-3) (3d-4) n -2 (d-3) n^2\bigr] {<}\tilde{O}_{n}{>}\\
&\qquad\qquad{} + n \bigl[d^3 - 5d^2 + 6d + 4 + (d-3) (d+4) n - 2 (d-3) n^2\bigr] {<}\tilde{O}_{n-1}{>}\Bigr)\\
&\qquad{} - n (n-1) (d-n+1) (2d-n-2) {<}\tilde{O}_{n-2}{>}\\
&\qquad{} + n (n-1) (n-2) (d-n+1) (d-n+2) {<}\tilde{O}_{n-3}{>}\biggr]\,.
\end{align*}
The matrix element ${<}\bar{b}d|O'_{n0}|b\bar{d}{>}$ is obtained by replacing
$C_F \to T_F (T_1 - T_2/N)$ in $M_{1,2}$, $T_F (T_2 - T_1/N) \to C_F$ in $M_{3,4}$,
and $T_F (T_2 - T_1/N) \to T_F (T_1 - T_2/N)$ in $M_5$.

\end{appendices}


\begin{thebibliography}{99}
\bibitem{Lenz:2006hd}
  A.~Lenz, U.~Nierste,
  JHEP {\bf 0706}, 072 (2007)
  [hep-ph/0612167].

\bibitem{Lenz} 
A.~Lenz,
Int.\ J.\ Mod.\ Phys.\ \textbf{A 23} (2008) 3321
[arXiv:0710.0940 [hep-ph]]

\bibitem{Nierste}
U.~Nierste,
arXiv:0904.1869 [hep-ph];
arXiv:1212.5805 [hep-ph].

\bibitem{Aoki:2016frl}
  S.~Aoki {\it et al.},
  arXiv:1607.00299 [hep-lat].

\bibitem{Aoki:2014nga}
  Y.~Aoki, T.~Ishikawa, T.~Izubuchi, C.~Lehner and A.~Soni,
  Phys.\ Rev.\ D {\bf 91} (2015) no.11,  114505
  [arXiv:1406.6192 [hep-lat]].

\bibitem{Grozin:2016uqy} 
  A.~G.~Grozin, R.~Klein, T.~Mannel and A.~A.~Pivovarov,
  Phys.\ Rev.\ D {\bf 94}, no. 3, 034024 (2016)
  [arXiv:1606.06054 [hep-ph]].

\bibitem{Kirk:2017juj} 
  M.~Kirk, A.~Lenz and T.~Rauh,
  JHEP {\bf 1712}, 068 (2017)
  [arXiv:1711.02100 [hep-ph]].

\bibitem{GL:09}
A.\,G.~Grozin, R.\,N.~Lee,
JHEP \textbf{02} (2009) 047
[arXiv:0812.4522 [hep-ph]].

\bibitem{Dowdall:2014qka} 
R.\,J.~Dowdall \textit{et al.} [HPQCD Collaboration],
PoS LATTICE \textbf{2014} (2014) 373
[arXiv:1411.6989 [hep-lat]].

\bibitem{Carrasco:2013zta}
  N.~Carrasco {\it et al.} [ETM Collaboration],
  JHEP {\bf 1403} (2014) 016
  [arXiv:1308.1851 [hep-lat]].

\bibitem{Bazavov:2016nty} 
  A.~Bazavov {\it et al.} [Fermilab Lattice and MILC Collaborations],
  Phys.\ Rev.\ D {\bf 93}, no. 11, 113016 (2016)
  [arXiv:1602.03560 [hep-lat]].

\bibitem{Buras:1990fn} 
  A.\,J.~Buras, M.~Jamin, P.\,H.~Weisz,
  Nucl.\ Phys.\ B {\bf 347}, 491 (1990).

\bibitem{BBL:96}
G.~Buchalla, A.\,J.~Buras, M.\,E.~Lautenbacher,
Rev.\ Mod.\ Phys.\ \textbf{68} (1996) 1125
[hep-ph/9512380].

\bibitem{Beneke:1998sy} 
  M.~Beneke, G.~Buchalla, C.~Greub, A.~Lenz, U.~Nierste,
  Phys.\ Lett.\ B {\bf 459}, 631 (1999)
  [hep-ph/9808385].

\bibitem{Inami:1980fz} 
  T.~Inami, C.\,S.~Lim,
  Prog.\ Theor.\ Phys.\  {\bf 65}, 297 (1981)
  Erratum: [Prog.\ Theor.\ Phys.\  {\bf 65}, 1772 (1981)].

\bibitem{Grozin:2017uto} 
  A.~G.~Grozin, T.~Mannel and A.~A.~Pivovarov,
  Phys.\ Rev.\ D {\bf 96}, no. 7, 074032 (2017)
  [arXiv:1706.05910 [hep-ph]].

\bibitem{tHooft:1972tcz} 
  G.~'t Hooft and M.~J.~G.~Veltman,
  Nucl.\ Phys.\ B {\bf 44}, 189 (1972).

\bibitem{Altarelli:1980fi} 
  G.~Altarelli, G.~Curci, G.~Martinelli and S.~Petrarca,
  Nucl.\ Phys.\ B {\bf 187}, 461 (1981).

\bibitem{Chanowitz:1979zu} 
  M.~S.~Chanowitz, M.~Furman and I.~Hinchliffe,
  Nucl.\ Phys.\ B {\bf 159}, 225 (1979).

\bibitem{Buras:1989xd} 
  A.~J.~Buras and P.~H.~Weisz,
  Nucl.\ Phys.\ B {\bf 333}, 66 (1990).

\bibitem{Dugan:1990df} 
  M.~J.~Dugan and B.~Grinstein,
  Phys.\ Lett.\ B {\bf 256}, 239 (1991).

\bibitem{Herrlich:1994kh}
  S.~Herrlich and U.~Nierste,
  Nucl.\ Phys.\ B {\bf 455} (1995) 39
  [hep-ph/9412375].

\bibitem{Pivovarov:1988gt} 
  A.\,A.~Pivovarov and L.\,R.~Surguladze,
  Sov.\ J.\ Nucl.\ Phys.\  {\bf 48}, 1117 (1989)
  [Yad.\ Fiz.\  {\bf 48}, 1856 (1988)].

\bibitem{Pivovarov:1991nk} 
  A.\,A.~Pivovarov and L.\,R.~Surguladze,
  Nucl.\ Phys.\ B {\bf 360}, 97 (1991).

\bibitem{Chetyrkin:1997gb} 
  K.~G.~Chetyrkin, M.~Misiak and M.~Munz,
  Nucl.\ Phys.\ B {\bf 520}, 279 (1998)
  [hep-ph/9711280].
 

\bibitem{N:94}
M.~Neubert,
Phys.\ Reports \textbf{254} (1994) 259.

\bibitem{MW:00}
A.\,V.~Manohar, M.\,B.~Wise,
\textit{Heavy Quark Physics},
Cambridge University Press (2000).

\bibitem{G:04}
A.\,G.~Grozin,
\textit{Heavy Quark Effective Theory},
Springer Tracts in Modern Physics \textbf{201},
Springer (2004).

\bibitem{Korner:2003zk}
  J.~G.~Korner, A.~I.~Onishchenko, A.~A.~Petrov and A.~A.~Pivovarov,
  Phys.\ Rev.\ Lett.\  {\bf 91}, 192002 (2003)
  [hep-ph/0306032].

\bibitem{CFG:96}
M.~Ciuchini, E.~Franco, V.~Gim\'enez,
Phys.\ Lett.\ \textbf{B 388} (1996) 167
[hep-ph/9608204].

\bibitem{FHH:91}
J.\,M.~Flynn, O.\,F.~Hern\'andez, B.\,R.~Hill,
Phys.\ Rev.\ \textbf{D 43} (1991) 3709.

\bibitem{B:96}
G.~Buchalla,
Phys.\ Lett.\ \textbf{B 395} (1997) 364
[hep-ph/9608232].

\bibitem{BG:95}
D.\,J.~Broadhurst, A.\,G.~Grozin,
Phys.\ Rev.\ \textbf{D 52} (1995) 4082
[hep-ph/9410240].

\bibitem{EH:90}
E.~Eichten, B.~Hill,
Phys.\ Lett.\ \textbf{B 234} (1990) 511.

\bibitem{Grozin:1998kf}
A.\,G.~Grozin,
Phys.\ Lett.\ \textbf{B 445} (1998) 165
[hep-ph/9810358].

\bibitem{Bekavac:2009zc}
S.~Bekavac, A.\,G.~Grozin, P.~Marquard, J.\,H.~Piclum, D.~Seidel and M.~Steinhauser,
Nucl.\ Phys.\ \textbf{B 833} (2010) 46
[arXiv:0911.3356 [hep-ph]].

\bibitem{Broadhurst:1991fy}
D.\,J.~Broadhurst, N.~Gray and K.~Schilcher,
Z.\ Phys.\ \textbf{C 52} (1991) 111.

\bibitem{Broadhurst:1991fi}
D.\,J.~Broadhurst,
Z.\ Phys.\ \textbf{C 54} (1992) 599.

\bibitem{Grozin:2012ec}
A.~Grozin,
Int.\ J.\ Mod.\ Phys.\ \textbf{A 28} (2013) 1350015
[arXiv:1212.5144 [hep-ph]].

\bibitem{PW:88}
H.\,D.~Politzer, M.\,B.~Wise,
Phys.\ Lett.\ \textbf{B 206} (1988) 681.

\bibitem{SV:88}
M.\,A.~Shifman, M.\,B.~Voloshin,
Sov.\ J.\ Nucl.\ Phys.\  \textbf{47} (1988) 511
[Yad.\ Fiz.\  \textbf{47} (1988) 801].

\bibitem{Grozin:2003gf}
  A.\,G.~Grozin,
  hep-ph/0311050.
 
\bibitem{Krasnikov:1996jq} 
  N.~V.~Krasnikov and A.~A.~Pivovarov,
  Mod.\ Phys.\ Lett.\ A {\bf 11}, 835 (1996)
  [hep-ph/9602272].

\bibitem{G:93}
V.~Gim\'enez,
Nucl.\ Phys.\ \textbf{B 401} (1993) 116.

\bibitem{Narison:1994zt} 
  S.~Narison, A.\,A.~Pivovarov,
  Phys.\ Lett.\ B {\bf 327}, 341 (1994)
  [hep-ph/9403225].

\bibitem{Ovchinnikov:1988zw}
  A.~A.~Ovchinnikov and A.~A.~Pivovarov,
  Phys.\ Lett.\ B {\bf 207} (1988) 333
   [Sov.\ J.\ Nucl.\ Phys.\  {\bf 48} (1988) 120]
   [Yad.\ Fiz.\  {\bf 48} (1988) 189].

\bibitem{MPP:11}
T.~Mannel, B.\,D.~Pecjak, A.\,A.~Pivovarov,
Eur.\ Phys.\ J.\ \textbf{C 71} (2011) 1607.

\bibitem{Chetyrkin:1985vj} 
  K.\,G.~Chetyrkin, A.\,L.~Kataev, A.\,B.~Krasulin, A.\,A.~Pivovarov,
  Phys.\ Lett.\ B {\bf 174}, 104 (1986)
  [hep-ph/0103230].

\bibitem{Penin:1998kx}
  A.~A.~Penin and A.~A.~Pivovarov,
  Nucl.\ Phys.\ B {\bf 549}, 217 (1999)
  [hep-ph/9807421].

\bibitem{Hoang:2000yr}
  A.~H.~Hoang {\it et al.},
  Eur.\ Phys.\ J.\ direct {\bf 2}, no. 1, 3 (2000)
  doi:10.1007/s1010500c0003
  [hep-ph/0001286].
  
\bibitem{Penin:1998wj}
  A.~A.~Penin and A.~A.~Pivovarov,
  Phys.\ Lett.\ B {\bf 443}, 264 (1998)
  doi:10.1016/S0370-2693(98)01323-9
  [hep-ph/9805344].

\bibitem{KM:92}
W.~Kilian, T.~Mannel,
Phys.\ Lett.\ \textbf{B 301} (1993) 382
[hep-ph/9211333].

\bibitem{Gelhausen:2013wia} 
  P.~Gelhausen, A.~Khodjamirian, A.~A.~Pivovarov, D.~Rosenthal,
  Phys.\ Rev.\ D {\bf 88}, 014015 (2013)
  Erratum: [Phys.\ Rev.\ D {\bf 89}, 099901 (2014)]
  Erratum: [Phys.\ Rev.\ D {\bf 91}, 099901 (2015)]
  [arXiv:1305.5432 [hep-ph]].

\bibitem{Rosner:2015wva} 
  J.~L.~Rosner, S.~Stone and R.~S.~Van de Water,
  [arXiv:1509.02220 [hep-ph]].

\bibitem{Ovchinnikov:1985bs} 
  A.~A.~Ovchinnikov and A.~A.~Pivovarov,
  Phys.\ Lett.\  {\bf 163B}, 231 (1985).

\bibitem{Korner:2000wd} 
  J.~G.~Korner, F.~Krajewski and A.~A.~Pivovarov,
  Eur.\ Phys.\ J.\ C {\bf 20}, 259 (2001)
  [hep-ph/0003165].

\bibitem{GSS:06}
A.\,G.~Grozin, A.\,V.~Smirnov, V.\,A.~Smirnov,
JHEP \textbf{0611} (2006) 022
[hep-ph/0609280].


\end{thebibliography}
\end{document}